\documentclass[prd,twocolumn,showpacs,reprint,preprintnumbers,nofootinbib]{revtex4-1}
\input{header}

\begin{document}

\title{Dark matter relic density from conformally or disformally coupled light scalars}

\author{Sebastian~Trojanowski}
\email{s.trojanowski@sheffield.ac.uk}
\affiliation{Consortium for Fundamental Physics, School of Mathematics and  Statistics, University of Sheffield, Hounsfield Road, Sheffield, S3 7RH, UK}
\affiliation{National Centre for Nuclear Research, Pasteura 7, 02-093 Warsaw, Poland}

\author{Philippe Brax}
\email{philippe.brax@ipht.fr}
\affiliation{Institut de Physique Th$\acute{\textrm{e}}$orique
, Universit$\acute{\textrm{e}}$ Paris-Saclay, CEA, CNRS, F-91191 Gif-sur-Yvette Cedex, France}

\author{Carsten van de Bruck}
\email{c.vandebruck@sheffield.ac.uk}
\affiliation{Consortium for Fundamental Physics, School of Mathematics and  Statistics, University of Sheffield, Hounsfield Road, Sheffield, S3 7RH, UK}

\begin{abstract}
Thermal freeze-out is a  prominent example of dark matter (DM) production mechanism in the early Universe  that can yield the correct relic density of stable weakly interacting massive particles (WIMPs). At the other end of the mass scale, many popular extensions of the Standard Model predict the existence of ultra-light scalar fields. These can be coupled to matter, preferentially in a universal and shift-symmetry-preserving way. We study the impact of such conformal and disformal couplings on the relic density of WIMPs, without introducing modifications to the thermal history of the Universe. This can either result in an additional thermal contribution to the DM relic density or suppress otherwise too large abundances compared to the observed levels. In this work, we assume that the WIMPs only interact with the standard model via the light scalar portal. We use simple models of fermionic or scalar DM, although a similar discussion holds for more sophisticated scenarios, and predict that their masses should be  between $\sim 100~\textrm{GeV}$ and several $\textrm{TeV}$ to comply both with the DM abundance and current bounds on the couplings of the light scalars to matter at the LHC. Future searches will tighten these bounds.
\end{abstract}

\renewcommand{\baselinestretch}{0.85}\normalsize
\maketitle
\tableofcontents
\renewcommand{\baselinestretch}{1.0}\normalsize

\section{\label{sec:intro}Introduction}

There is an overwhelming amount of evidence in favour of  the gravitational existence of  dark matter (DM)  and its impact on baryons. Understanding the nature and origin of DM remains an open issue due  to the null searches for its couplings to the Standard Model (SM) at both accelerators and astrophysically. Although purely gravitational production of DM is possible (see e.g.~\cite{Ema:2015dka,Garny:2015sjg,Tang:2016vch} and references therein), it is often sensitive to the assumed initial conditions in the early Universe, and it is relevant only in the absence of additional interactions that could equilibrate DM with the thermal plasma at a later stage~\cite{Tang:2017hvq}.

On the other hand, the common scenario in which DM undergoes a period of thermal equilibrium with the SM species is an exceptionally attractive production mechanism. In this case, the DM relic density, $\Omega_\chi h^2$,  freezes out and is essentially insensitive to the evolution of the Universe prior to the decoupling time. Instead, $\Omega_\chi h^2$ is governed by an interplay between the DM mass and the coupling strength to the SM at the freeze-out temperature. This scenario has given rise to a plethora of models predicting cold DM (CDM) consisting of Weakly Interacting Massive Particles (WIMPs), including e.g. supersymmetry (see~\cite{Roszkowski:2017nbc} for a recent review) but also a number of effective models in which DM is coupled to the SM through simple portals, cf. Ref.~\cite{Arcadi:2017kky}.

However, the present null results of direct and indirect DM searches put ever-tightening bounds on the couplings of DM to the SM. In some of the simplest cases, the thermal value of the WIMP-like DM annihilation cross section is already excluded for masses below $\sim 100~\textrm{GeV}$~\cite{Ackermann:2015zua}, while similar bounds from the Cherenkov Telescope Array~\cite{Acharya:2017ttl} will reach the same level in the coming years for even larger masses, up to $m_\chi\sim \textrm{several TeV}$, although this may depend on the dominant annihilation final state and the actual DM profile towards the Galactic Center (GC)~\cite{Hryczuk:2019nql}. The absence of any DM discovery in the near future could then point towards suppressed couplings to the SM. Unfortunately this often results in a freeze out of DM particles which happens too early in the early Universe so that then they become overproduced.

At the other end of the mass scale, ultra-light scalar fields are another type of popular species of particles beyond the SM (BSM). They appear in models related to dynamical dark energy~\cite{Copeland:2006wr} driven by quintessence~\cite{Amendola:1999er} or emerge from modified gravity theories~\cite{Clifton:2011jh}, but also e.g. in scenarios with fuzzy DM~\cite{Hu:2000ke}. In addition, the appearance of at least one such ``Stuckelberg'' field could play an essential role in restoring the diffeomorphism invariance of the action in the expanding Universe~\cite{Gubitosi:2012hu}. It is then natural to assume that at least some of such ultra-light scalars will couple to both gravity and matter and even possibly drive the accelerating expansion of the Universe~\cite{Joyce:2014kja}. Such couplings are expected to be universal to all matter species, including the SM and DM, as dictated by the weak equivalence principle, although a violation of this rule is also allowed to some extent by current observations that constrain more strongly interactions with baryons~\cite{Adelberger:2003zx}.

The mass of such scalars $\phi$ could be  low and should  be protected from possible large radiative corrections induced by $\phi$ couplings to matter thanks to a shift symmetry. A natural scenario employs the shift symmetry, $\phi\rightarrow\phi+c$, where $c$ is constant, which is preserved by derivative couplings of $\phi$. One expects this symmetry to be only mildly broken by a non-zero mass of $\phi$. Hence interaction terms with unsuppressed couplings, which would explicitly break the shift symmetry, are assumed to be absent.

In this study, we analyze how the thermal production of WIMP DM in the early Universe is inevitably affected by the presence of such ultra-light scalar fields coupled to both the SM and DM. To this end, we employ simple conformal and disformal shift-symmetry-preserving operators. We show that $\phi$-mediated couplings alone can easily lead to the observed value of DM relic abundance in the regions of the parameter space of the models that are currently not excluded, and  might be further testable in future searches at the Large Hadron Collider (LHC). In particular, this could provide a mechanism for models predicting  too weak DM couplings to the SM and, therefore, too large abundances of WIMP DM, to be compatible with data eventually. As discussed above, such new scenarios with new scalar couplings  might become of relevance in the light of current and future observational bounds.

The particular simple interaction terms that we analyze can naturally appear in the context of modified gravity, as briefly described below, but can also be considered at a more general level. We limit ourselves to the dominant operators that are normally screened from detection prospects at low energies, relevant for most terrestrial experiments and fifth force searches. On the other hand, they effectively switch on at a high energy scale corresponding to the LHC and to the early Universe when WIMP DM production occurred. 

This paper is organized as follows. In \cref{sec:theory}, we introduce the conformal and disformal interactions of light scalars $\phi$ with matter and discuss relevant constraints, as well as basic limitations of the effective field theory approach used to study them. In \cref{sec:reldens}, we study WIMP DM relic density emerging from the $\phi$-portal to the SM, while in \cref{sec:pheno} we discuss its phenomenological impact on the LHC, DM, and gravitational-wave searches. We conclude in \cref{sec:conclusions}. Selected more technical details are given in appendices. In particular, in \cref{app:otherinteractions} we discuss additional interactions that can appear in models under study, but typically play a subdominant role in determination of the DM relic density. In \cref{app:Boltzmann,app:SE} we provide more details about solving the relevant Boltzmann equations and about the Sommerfeld enhancement factor for DM annihilations coupled to the SM via such a $\phi$-portal, respectively. In \cref{app:scattering} we discuss the scattering cross sections of disformally and conformally coupled scalars. \Cref{app:EMgravwaves} is devoted to discussion about a difference between gravitational and electromagnetic waves propagation in a disformally modified metric.

\section{\label{sec:theory}Conformal and disformal couplings of ultra-light scalars}

As described above, light scalar degrees of freedom are amongst the most often discussed extensions of the SM, and they play a very special role in cosmology. A prominent example are ultra light scalar fields $\phi$ appearing in many dark energy models~\cite{Copeland:2006wr}, which can be non-trivially coupled to matter~\cite{Clifton:2011jh,Brax:2017idh}. Such interactions can naturally arise from a coupling of the $\phi$ field to the space-time metric as in scalar-tensor theories. At leading order, this affects the motion of matter fields and modifies geodesics, which can become $\phi$ dependent.

\subsection{Light scalars and  modified gravity\label{sec:modgrav}}

The most general of such coupling preserving Lorentz invariance and causality can be written~\cite{Bekenstein:1992pj} using the following transformation between the metric $g_{\mu\nu}$ in the Einstein frame, in which the gravitational part of the theory is the standard Einstein-Hilbert action of General Relativity, identified below with the Minkowski metric $g_{\mu\nu}=\eta_{\mu\nu}$ and the one in the Jordan frame, where particles are canonically coupled to ${\tilde g}_{\mu\nu}$,
\begin{equation}
\tilde{g}_{\mu\nu} = C(\phi,X)\,g_{\mu\nu} + D(\phi,X)\,\partial_\mu\phi\,\partial_\nu\phi.
\label{eq:metric}
\end{equation}
The first term in \cref{eq:metric} is the  conformal coupling while the second one proportional to $D(\phi,X)$ is referred to as disformal. The functions $C$ and $D$ can depend on the field $\phi$ and on the kinetic term $X = -(1/2)\,g^{\mu\nu}\,\partial_\mu\phi\,\partial_\nu\phi$.

Conformal metric transformations naturally arise in the context of scalar-tensor theories~\cite{Damour:1992we,Faraoni:1998qx} and have been thoroughly investigated in the literature (see~\cite{Brax:2017idh} for recent review). In particular, such theories involving kinetically dependent conformal couplings have been investigated in Ref.~\cite{Brax:2016kin}. On the other hand, the additional disformal terms can appear in Horndeski-type~\cite{Horndeski:1974wa} theories~\cite{Bettoni:2013diz,Zumalacarregui:2013pma}, D-brane scenarios~\cite{deRham:2010eu,Koivisto:2013fta}, branon models~\cite{Dobado:2000gr,Alcaraz:2002iu,Cembranos:2004jp} or in the context of massive gravity~\cite{deRham:2010ik,deRham:2010kj}. For simplicity, one often considers a purely conformal scenario, in which $D= 0$, or a purely disformal one with $C= 1$, although both types of couplings can simultaneously appear in specific models.

Light scalar fields coupled to matter can induce  long-range fifth forces~\cite{Carroll:1998zi,Amendola:1999er} which are  strongly constrained for baryons~\cite{Adelberger:2003zx}, unless  screened from local tests of gravity~\cite{Joyce:2014kja}. A fifth force in the DM sector~\cite{Damour:1990tw,Friedman:1991dj,Farrar:2003uw} can also be constrained by  observations of the  anisotropies of  the cosmic microwave background (CMB) radiation, large scale structures and supernovae data~\cite{Amendola:2003eq,Bean:2008ac,Xia:2009zzb,Amendola:2011ie,Pettorino:2012ts,Pettorino:2013oxa,Bolotin:2013jpa,vandeBruck:2015ida,vandeBruck:2016hpz,Mifsud:2017fsy,Miranda:2017rdk,vandeBruck:2017idm,Vagnozzi:2019kvw}. A  large discrepancy between the fifth forces in the SM and DM sectors could also lead to observable effects related to the baryon bias~\cite{Amendola:2001rc,Kesden:2006zb}. On the other hand, non-negligible late-time interactions in the DM sector might avoid current bounds on the couplings and could be favored by cosmological data~\cite{Abdalla:2014cla,DiValentino:2019jae}. This could point towards  interactions between the DM and a quintessential scalar field with a strength which grows in time, see e.g. \cite{vandeBruck:2019vzd}.

At the  microscopic level, interactions of $\phi$ with matter can be deduced from a series expansion of the action around the vacuum characterized by $\phi=0$ and the metric $g_{\mu\nu}=\eta_{\mu\nu}$~\cite{Kugo:1999mf,Kaloper:2003yf}. The actual form of the interaction terms  depend on the expansion around $\phi,X\approx 0$ of the functions $C$ and $D$. In particular, for $C(X) \approx 1+X/M^4$ one obtains the conformal interaction Lagrangian discussed below~\cite{Brax:2016kin}, while the disformal one can be derived from e.g. exponential function $D  = (2/M^4)\,e^{\beta_{i}\phi} \simeq 2/M^4$.

In the following, we will discuss the phenomenology of simple conformal and disformal portals between the SM and DM. While, in general, our  considerations are independent of the specific theoretical motivation that lies behind these portals, it is useful to keep in mind their possible connections to other  problems in contemporary physics, including the nature of the dark energy content of the Universe.

\subsection{\label{sec:model}Models}

In order to avoid model-dependent issues related to various possible UV completions of the models under study, we will follow an effective field theory (EFT) approach~\cite{Brax:2016did} assuming that the effects related to heavy new physics can be integrated out at the energies relevant for our discussion. We focus on the models that couple universally to all the SM species and preserve a shift symmetry of the $\phi$ field. For simplicity, we also assume that the DM particle $\chi$ is a Dirac fermion
\begin{equation}
\mathcal{L}_{\textrm{DM,ferm.}} = i\bar{\chi}\gamma^\mu\partial_\mu\chi - m_\chi\bar{\chi}\chi,
\label{eq:LDM}
\end{equation}
although other possibilities could also lead to similar phenomenology, and we will comment below on the case of disformally coupled scalar DM,
\begin{equation}
\mathcal{L}_{\textrm{DM,c.scal}} = (\partial_\mu\chi)^\dagger(\partial^\mu\chi) - m^2_\chi\chi^\dagger\chi.
\label{eq:LDMcompscal}
\end{equation}

When working within the framework of theories of modified gravity, $\chi$ DM particle moving along the geodesics corresponding to the metric given in \cref{eq:metric} will gain $\phi$-dependent masses. However, for small modifications introduced to the Minkowski metric, the value of $m_\chi$ can be assumed constant at leading order. On the other hand, at  the next-to-leading order in the $m_\chi(\phi)$ expansion, non-trivial couplings between $\chi$ and $\phi$ appear. We will discuss example of such interactions below, writing down effective Lagrangians that could also be studied independently of the modified-gravity context.

\begin{figure}[tb]
\includegraphics[width=0.4\textwidth]{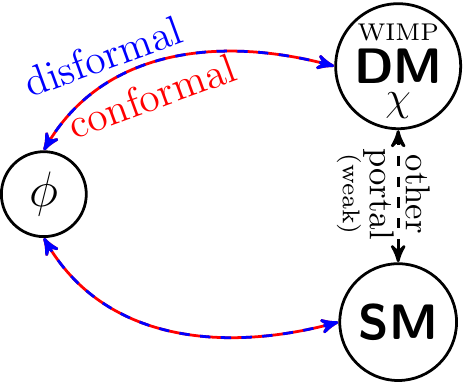}
\caption{A schematic illustration of a conformally and disformally coupled scalar $\phi$ acting as a portal between the SM and dark matter (DM) consisting of weakly-interacting massive particles (WIMPs) $\chi$. Other direct couplings to the SM can also contribute to the DM relic density, while the impact of intermediate $\phi$s remains important. In this study, we assume that the other portals between the SM and DM are weaker and play a subdominant role.}
\label{fig:whatsgoingon}
\end{figure}

The DM $\chi$ particle can also have other couplings to the SM, as schematically illustrated in \cref{fig:whatsgoingon}. In the following, we assume that these are subdominant, which allows us to focus on the sole impact of the $\phi$-mediated interactions. In particular, this corresponds to the case, in which $\chi$ particles would never be in thermal equilibrium in the early Universe in the absence of the additional coupling to $\phi$, or to the scenario in which the $\chi$ particles would equilibrate but freeze-out too early with a large relic abundance. We leave for future studies  the analysis of a possible interplay between the SM-DM portals considered below and other types of couplings if both are tuned to play a comparable role. It is useful, however, to stress that even if other interactions of $\chi$ DM could cause its freeze-out with $\Omega_\chi h^2\sim 0.1$, the impact of the coupling to $\phi$ cannot be easily dismissed from the discussion about DM relic density. Instead, as discussed below, the contribution from interactions with $\phi$ can easily be sizeable around the typical freeze-out temperatures of WIMP-like DM, $T_{\chi,\textrm{fo}}\sim m_\chi/20$.

The dominant shift-symmetry-preserving interactions of $\phi$ with the SM and DM correspond to the conformal derivative coupling
\begin{equation}
\mathcal{L}_{\textrm{C},\phi-\textrm{SM/DM}}=\frac{\partial_\mu\phi\,\partial^\mu\phi}{M^4_{\textrm{C,SM/DM}}}\,(T_{\textrm{SM/DM}})_{\nu}^{\nu},
\label{eq:Lphiconf}
\end{equation}
and the disformal one
\begin{equation}
\mathcal{L}_{\textrm{D},\phi-\textrm{SM/DM}}=\frac{\partial_\mu\phi\,\partial_\nu\phi}{M^4_{\textrm{D,SM/DM}}}\,T_{\textrm{SM/DM}}^{\mu\nu},
\label{eq:Lphidisf}
\end{equation}
where we have introduced effective conformal and disformal mass scales $M_{\textrm{SM}}$ ($M_{\textrm{DM}}$) that characterize the $\phi$ interactions with the SM (DM). Given our phenomenological approach, we allow the relevant scales for the SM and DM to be set independently, $M_{\textrm{SM}}\neq M_{\textrm{DM}}$, although our main interest is in the case when they are comparable, $M_{\textrm{SM}}\sim M_{\textrm{DM}}$. Such universal couplings could be additionally motivated by their possible common origin. 

The $\phi$ field couples to the usual energy-momentum tensor of the SM and DM particles. In particular, the relevant expression for fermionic $\chi$, which neglects terms vanishing for on-shell fermions, reads\footnote{Note that an additional factor of $1/2$ appears in the symmetrization.}
\begin{equation}
T_{\textrm{DM,ferm}}^{\mu\nu} = \frac{i}{2}\left[\bar{\chi}\gamma^{(\mu}\partial^{\nu)}\chi - \partial^{(\mu}\bar{\chi}\gamma^{\nu)}\chi\right],
\label{eq:Tmunu}
\end{equation}
where on the right hand side the indices have been symmetrized. In case of the SM, similar equation holds but with the covariant derivative, $\partial\rightarrow D$, which also determines the interactions between $\phi$ and the SM gauge bosons.

The energy momentum tensor for a complex scalar field $\chi$ is given by
\begin{align}
T_{\textrm{DM,c.scal.}}^{\mu\nu} =& -g^{\mu\nu}\left[(\partial_\rho\chi)^\dagger(\partial^\rho\chi)-m_\chi^2\,\chi^\dagger\chi\right]\nonumber\\
 &+ (\partial^\mu\chi)^\dagger(\partial^\nu\chi)+ (\partial^\nu\chi)^\dagger(\partial^\mu\chi).
\label{eq:Tmunucomplscal}
\end{align}
On top of direct couplings to $\phi$, additional interactions between the SM and DM particles can occur at the loop-level. However, these typically play a subdominant role when analyzing the $\chi$ relic density, similarly to possible scalar self-couplings, as discussed in~\cref{app:otherinteractions}.

Notably, the models described by \cref{eq:Lphiconf,eq:Lphidisf,eq:Tmunu,eq:Tmunucomplscal} couple pairs of $\phi$ or $\chi$ fields, therefore allowing to introduce additional discrete symmetries stabilizing the relevant particles on cosmological time scales. 

\subsection{\label{sec:interactions}Dominant interaction cross sections}

The most relevant interaction cross sections for our discussion concern annihilation processes of a pair of Dirac fermions into a pair of light scalars, $\bar{f}f\rightarrow\phi\phi$, where $f$ can correspond to either the SM or DM particles.\footnote{In case of disformal couplings to the SM, an important role is also played by annihilations involving gluons, cf. \cref{eq:gluons}.} In particular, assuming a negligible scalar mass, $m_\phi\ll m_\chi$, the cross section for fermionic DM annihilation reads
\begin{equation}
\sigma_{\textrm{ann,C,ferm.}} = \frac{m_\chi^2\,s^2}{256\pi\,M^8_{\textrm{DM,C}}}\,\sqrt{1-\frac{4\,m_\chi^2}{s}},
\label{eq:sigannconf}
\end{equation}
for the conformal model, where $s$ is a square of the center-of-mass energy, and~\cite{Cembranos:2003fu,Cembranos:2016jun}
\begin{equation}
\sigma_{\textrm{ann,D,ferm.}} = \frac{1}{30720\,\pi}\,\frac{s^3}{M_{\textrm{DM,D}}^8}\,\sqrt{1-\frac{4m_\chi^2}{s}}\,\left[1+6\frac{m_\chi^2}{s}\right],
\label{eq:siganndisf}
\end{equation}
for the disformal one. Similar expressions hold for the SM fermions with $M_{\textrm{DM}}$ replaced with $M_{\textrm{SM}}$. In particular, for both models, the annihilation cross section vanishes in the non-relativistic limit, $s\simeq 4\,m_\chi^2$, with the dominant p-wave contribution equal to
\begin{equation}
\sigma_{\textrm{ann,ferm.}}v \stackrel{v\ll 1}{\simeq} \frac{m_\chi^6}{A\pi\,M_{\textrm{DM}}^8}\,v^2, 
\label{eq:sigmaannpwave}
\end{equation}
where $A = 32$ ($384$) for the conformal (disformal) case, and $v$ is the relative velocity between the DM particles. As can be seen, the thermal value of the annihilation cross section, $\sigma_{ann}v\sim \textrm{a few} \times 10^{-9}\,\textrm{GeV}^{-2}$, is obtained for a typical ratio $m_\chi^6/M_{\textrm{DM}}^8\sim (10^{-6}-10^{-5})~\textrm{GeV}^{-2}$ assuming a characteristic WIMP velocity at freeze-out, $v^2\sim 0.1$. This ratio can naturally be obtained for $m_\chi\sim M_{\textrm{DM}}\sim (0.1-\textrm{a few})~\textrm{TeV}$. We will discuss the DM relic density in this scenario below in \cref{sec:reldens,app:Boltzmann}. Notably, at present times $\chi$ DM annihilations into $\phi$ particles in the GC are negligible, given typical velocities of order $v\sim 10^{-3}$ and only very mild expected impact of the Sommerfeld enhancement, cf. \cref{app:SE}. 

In the discussion below, we also refer to the model with disformally coupled scalar DM $\chi$. The relevant annihilation cross section reads~\cite{Cembranos:2003fu,Cembranos:2016jun}
\begin{equation}
\sigma_{ann,\textrm{D,c.scal.}} = \frac{1}{3840\,\pi}\,\frac{s^3}{M_{\textrm{DM,D}}^8}\,\frac{\left[1+\frac{2\,m_\chi^2}{s}+\frac{6\,m_\chi^4}{s^2}\right]}{\sqrt{1-\frac{4m_\chi^2}{s}}},
\label{eq:siganndisfscalar}
\end{equation}
which, in the non-relativistic regime, is dominated by the s-wave contribution
\begin{equation}
\sigma_{\textrm{ann,D,c.scal.}}\,v\stackrel{v\ll 1}{\simeq} \frac{m_\chi^6}{16\pi\,M_{\textrm{DM}}^8}.
\label{eq:sigmaannswave}
\end{equation}
This cross section is larger by a factor of a few or so from $\sigma_{\textrm{ann}} v$ obtained for both previous models, cf. \cref{eq:sigmaannpwave}. Because of this, as well due to lack of velocity suppression, the thermal value of $\sigma_{\textrm{ann}} v$ in this model can be obtained for slightly larger values of the $M_{\textrm{DM}}$ parameter. Still, however, the correct DM relic density is obtained along the lines of constant ratio $m_\chi^6/M_{\textrm{DM}}^8\sim 10^{-6}~\textrm{GeV}^{-2}$.

\subsection{\label{sec:bounds}Current bounds and validity of EFT approach}

\paragraph{Collider searches} The microscopic description of conformally and disformally coupled $\phi$-mediated interactions between baryons can be used to derive constraints on such couplings based on various astrophysical probes and terrestrial experiments (see e.g.~\cite{Brax:2014vva}). A prominent role in these efforts is played by collider searches, especially at the LHC~\cite{Brax:2015hma,Brax:2016did}. The most stringent such bounds based on $13~\textrm{TeV}$ LHC data have recently been published by the ATLAS collaboration~\cite{Aaboud:2019yqu}.
In particular, the results of the $t\bar{t}\!+\!E_T^{\textrm{miss}}$ and $\textrm{jet}\!+\!E_T^{\textrm{miss}}$ searches have been reinterpreted as limits on $M_{\textrm{SM}}$ for conformal (cf.~\cref{eq:Lphiconf})
\begin{equation}
M_{\textrm{C,SM}} \gtrsim 200~\textrm{GeV},
\label{eq:MSMLHCconf}
\end{equation}
and disformal (cf.~\cref{eq:Lphidisf})
\begin{equation}
M_{\textrm{D,SM}} \gtrsim 1200~\textrm{GeV},
\label{eq:MSMLHCdisf}
\end{equation}
models. The relative weakness of the bounds for the conformal model emerge from the dominant coupling of $\phi$ to heavy top quarks and experimental challenges in performing the $t\bar{t}\!+\!E_T^{\textrm{miss}}$ search.

\paragraph{Validity of EFT in the SM sector} These LHC constraints have been derived assuming that the validity of the EFT approach and unitarity hold in the relevant processes. In the analyses, only limited values of the centre-of-mass energy of the hard interaction have been allowed, $\sqrt{\hat{s}}<g_{\ast}\,M_{\textrm{SM}}$, where $g_{\ast}$ parameterizes the impact of an unknown UV completion that should not be omitted for too large momentum transfer~\cite{Berlin:2014cfa,Racco:2015dxa}. In particular, when selecting events, a simplified iterative procedure to recast bounds on $M_{\textrm{SM}}$ has been applied following Ref.~\cite{Busoni:2014sya}. This is relevant for a cut-and-count analysis, while more sophisticated studies would require introducing cuts at a generator level~\cite{Abercrombie:2015wmb}. In the following, we adopt the effective bounds from \cref{eq:MSMLHCconf,eq:MSMLHCdisf} that are relevant for $g_{\ast}\gtrsim \pi^2$ or $g_{\ast}\gtrsim \pi$ for conformal and disformal couplings, respectively~\cite{Aaboud:2019yqu}.

We note that if $g_\ast$ is to be interpreted as some effective mediator coupling strength, which corresponds to the UV complete theory turning on at some high energy scale, then the aforementioned values of this parameter lie in the strongly-coupled regime. Still, however, there is a room for them not to violate the perturbativity limit, $g_\ast<4\pi$, which we assume to be the case in the following.

A detailed treatment of a unitarity constraint based on a partial wave expansion of the relevant amplitudes could impose additional bounds on the cut-off scales discussed above, depending on the characteristic energy of the process (see e.g. Ref~\cite{Abercrombie:2015wmb} and references therein). In order to avoid these complications, we will limit our considerations to processes involving centre-of-mass energies $\sqrt{s}<\mathcal{O}(10~\textrm{TeV})$ characteristic for the LHC, where we assume that the EFT approach remains valid, while energies relevant for DM production will be even significantly lower.

\paragraph{Bounds on the coupling to DM} 
Direct microscopic coupling between DM and $\phi$ remains less constrained~\cite{Simpson:2010vh}. However, in presence of sizeable such interactions in the dark sector, the fifth force between baryons could be induced at a loop level~\cite{Bovy:2008gh,Carroll:2008ub}, which is tightly bounded, as discussed above. In addition, if shift-symmetry-breaking interaction terms are present, the DM-$\phi$ coupling could be further constrained by studying radiative corrections to the $\phi$ potential and its mass~\cite{DAmico:2016jbm}. We note, however, that by imposing additional discrete symmetries allowed in the interaction Lagrangians \cref{eq:Lphiconf,eq:Lphidisf}, the models of  interest can be further protected from inducing at  radiative level such shift-symmetry-breaking terms that would involve an odd number of  $\phi$ fields.

In the following, we then focus on shift-symmetry-preserving conformal and disformal models with an ultralight scalar field $\phi$ and a relatively heavy DM particle, $m_{\chi}\sim \textrm{TeV}$. Applying the EFT approach to interactions between $\chi$ and $\phi$ in these scenarios, including a particle production process in the early Universe, constrains the $M_{\textrm{DM}}$ parameter to large values, of a similar order to $M_{\textrm{SM}}$. This suppresses any possible long-range fifth force between DM particles to be beyond current observational capabilities.

\paragraph{Validity of EFT in the DM sector} In order to ensure the validity of the EFT approach in the DM-$\phi$ coupling, we perform a simple self-consistency check. In particular, as discussed in \cref{sec:thersigmavTfo}, the typical freeze-out temperatures of $\chi$-$\phi$ interactions are relatively low, of order $T_{\chi,\textrm{fo}}\sim m_\chi/20$, similarly to standard WIMP DM interactions. The $\chi$ DM particles then decouple being non-relativistic, and the relevant momentum transfer in the annihilation process $\chi\bar{\chi}\rightarrow\phi\phi$ is dominated by the mass of $\chi$. Similarly to the discussion above for the interactions between the SM and $\phi$, we require $\sqrt{s_{\textrm{DM}}}< g_{\ast}\,M_{\textrm{DM}}$, where $\sqrt{s_{\textrm{DM}}}$ is the center-of-mass energy of DM annihilations. In particular, around freeze-out one obtains $\sqrt{s_{\textrm{DM}}}\simeq 2\,m_\chi$, and the corresponding bound reads $M_{\textrm{DM}}\gtrsim c\,m_\chi$, where $c$ depends on the value of $g_{\ast}$. For the values of $g_{\ast}$ larger than the aforementioned limits used in the ATLAS study~\cite{Aaboud:2019yqu}, but lower than $4\pi$, one obtains $0.1 \lesssim c\lesssim 1$. For simplicity, in the following, we require
\begin{equation}
M_{\textrm{DM}}\gtrsim m_\chi,
\label{eq:MDMlimit}
\end{equation}
when identifying the regime in which EFT approach remains valid, while the effects of the UV completion can be neglected.

\paragraph{Other constraints relevant for ultra-high energy} For large energies, additional bounds on the models under study are related to ultra high-energy (UHE) cosmic-ray and astrophysical probes. In particular, as shown in \cref{app:scattering}, the scattering cross section of conformally or disformally coupled scalar off SM fermion can easily grow to large values for increasing energy of the incident scalar, $E_\phi$. In the rest frame of  protons, the centre-of-mass energy of such collision such that $\sqrt{s}\sim \mathcal{O}(10~\textrm{TeV})$, corresponds to $E_\phi\sim (10-100)~\textrm{PeV}$. The relevant scattering cross section can be as large as $\textrm{nb}-\mu\textrm{b}$, i.e., it can exceed the scattering cross sections of neutrinos at these energies, $\sigma_{\nu p}\sim \textrm{nb}$~\cite{Formaggio:2013kya}, by even $3$ orders of magnitude. The prospects of probing such scalars produced in UHE cosmic-ray showers in the atmosphere and subsequently scattering in neutrino telescopes are, however, degraded by a small production cross section, e.g. $\sigma(pp\rightarrow\phi\phi+X)\lesssim \textrm{pb}$ for $\sqrt{s}\sim 10~\textrm{TeV}$~\cite{Brax:2016did}.

On the other hand, both the production and scattering cross sections could be increased for even larger energy. This would, however, potentially violate EFT validity conditions and, therefore, the results of such an analysis would then be more sensitive to details of an unknown UV completion of the model.

\paragraph{Cosmological bounds on ultra-light scalars}
Ultra-light scalars conformally or disformally coupled to the SM at temperatures around the neutrino decoupling and the onset of the Big Bang Nucleosynthesis (BBN), $T_{\textrm{BBN}}\sim \textrm{MeV}$, could contribute to the effective number of relativistic degrees of freedom and therefore change the expansion history of the Universe. This could affect successful BBN or CMB predictions. This effect is typically parameterized in terms of the number of effective neutrino-like species, $N_{\textrm{eff}}$, which is bounded from above, $\Delta N_{\textrm{eff}} <0.285$, by the Planck data combined with the measurements of baryonic acoustic oscillations (BAO)~\cite{Aghanim:2018eyx}. 

However, as shown in \cref{sec:thersigmavTfo}, for $M_{\textrm{SM}}\gtrsim 100~\textrm{GeV}$, the decoupling of $\phi$ from the SM typically occurs at larger temperatures, $T_{\phi,\textrm{fo}}\sim \textrm{GeV}$ or so. As a result, the subsequent heating of the SM thermal bath around the time of the QCD phase transition generates a difference between the SM sector temperature $T$ and the dark sector temperature of the light scalar $\phi$. The corresponding contribution from $\phi$ to $N_\nu$ at the time of BBN can then be deduced from the conservation of entropy. It is given by (see e.g. Ref.~\cite{Cembranos:2016jun})
\begin{equation}
\Delta N_{\textrm{eff}} = \frac{4}{7}\,\left(\frac{10.75}{g_{\ast}(T_{\phi,\textrm{fo}})}\right)^{4/3}, 
\label{eq:DelNeff}
\end{equation}
where the number of relativistic degrees of freedom at the temperature of $\phi$ decoupling $T_{\phi,\textrm{fo}}$ is denoted by $g_{\ast}(T_{\phi,\textrm{fo}})$, while at the time of BBN $g_{\ast}(T_{\textrm{BBN}}) = 10.75$. Given that $g_{\ast}(T_{\phi,\textrm{fo}})\gtrsim 60$ before the QCD phase transition, one obtains $\Delta N_{\textrm{eff}}\lesssim 0.06$, which is well below the aforementioned current upper limit.

Once produced in the early Universe, stable light scalar particles that decouple whilst being relativistic could contribute to a hot DM (HDM) component, on top of a similar such contribution from relic neutrinos. A detailed analysis of the allowed value of $\Omega_\phi h^2$ would require marginalizing over the unknown SM neutrino masses, as well as taking into account different decoupling time and particle mass of light scalars. A simple estimate based on the upper limit on the abundance of relic neutrinos, $\Omega_\nu h^2\lesssim 0.0014$~\cite{Tanabashi:2018oca}, leads to an upper bound on the light scalar mass $m_\phi\lesssim \mathcal{O}(\textrm{eV})$, cf. \cref{eq:Oh2relativistic} below. In the following, we will assume that $\phi$ particles are lighter than this upper limit. 

\section{Relic density of conformally and disformally coupled WIMP dark matter\label{sec:reldens}}

As discussed above, the annihilation cross sections between the SM and light scalars, as well as between $\phi$ and DM, can be large enough so that both  BSM species remain in equilibrium with the thermal bath in the early Universe. Once the temperature $T$ drops down, the stable dark sector particles decouple from thermal plasma and \textsl{freeze-out} with some relic abundance, $\Omega_\chi h^2$ and $\Omega_\phi h^2$, that can be obtained by solving the relevant Boltzmann equations.

Before we discuss this in more detail, it is important to note that the presence of light quintessence-like scalar fields could affect the evolution of the Universe by, e.g. introducing an early kination phase. This would then lead to an additional effect that could further modify the relic abundance of DM~\cite{Kamionkowski:1990ni,Salati:2002md,Rosati:2003yw,Profumo:2003hq}. This scenario in the context of scalar-tensor theories of gravity has been first discussed in Refs~\cite{Catena:2004ba,Catena:2009tm} (see also Ref.~\cite{Dutta:2016htz} for further discussion) where a possible overall enhancement of $\Omega_\chi h^2$ of about several orders of magnitude has been reported. Beside such a strong impact on the expansion rate at early times, the subsequent evolution of the Universe would closely resemble general relativity due to the attraction mechanism which drives the coupling between the scalar field and metric to a constant value~\cite{Damour:1992kf,Damour:1993id,Bartolo:1999sq}. In addition, a rapid transition to the standard cosmological scenario could further affect the DM relic abundance by initiating reannihilation processes~\cite{Catena:2004ba}. The impact on the DM relic abundance, however, depends crucially on the coupling strength that modifies the metric. In particular, as discussed e.g. in Refs~\cite{Damour:1998ae,Coc:2006rt,Meehan:2015cna}, stringent bounds from the Big Bang Nucleosynthesis (BBN) can introduce important constraints on such scenarios that also limit the impact on $\Omega_\chi\,h^2$. 

In particular, the specific case of disformal coupling has recently been studied in Ref.~\cite{Dutta:2017fcn}, where the enhancement in the expansion rate was shown to reach up to a factor $\mathcal{O}(100)$. This, however, depends on the assumed initial conditions for the scalar field evolution, as well as on the value of $M_{\textrm{SM}}\sim M_{\textrm{DM}}$, and the resulting impact can be much smaller and shifted towards large temperatures above the one corresponding to $\chi$ freeze-out. We will neglect such possible modifications to the Hubble expansion rate in the following and assume that the energy budget of the Universe around the time of $\chi$ freeze-out is radiation dominated. We note, however, that, by carefully choosing the aforementioned initial conditions, $\Omega_\chi h^2$ could be further changed, on top of the thermal equilibrium effects discussed below.

As discussed in more details in \cref{app:Boltzmann}, we are interested in a typical scenario  where the $\chi$ DM freeze-out, characterized by the temperature $T_{\chi,\textrm{f.o}}$,  does not exceed the temperature of the electroweak phase transition, $T_{\textrm{EW}}$. On the other hand, the decoupling of light scalars from the SM thermal plasma occurs at a  lower temperature, $T_{\phi,\textrm{f.o}}$, although  before the QCD phase transition, $T_{\textrm{QCD}}$. Hence we consider $T_{\textrm{EW}}>T_{\chi,\textrm{f.o}}>T_{\phi,\textrm{f.o}}>T_{\textrm{QCD}}$. The last condition allows one to avoid complications due to the presence of hadronized particles among the final states of the $\phi\phi$ annihilations. This also helps avoiding  troublesome bounds from BBN, cf. \cref{sec:bounds}.

\subsection{Boltzmann equations} 

The evolution of the number densities in the system composed of two dark species $\phi$ and $\chi/\bar{\chi}$ out of which $\phi$ is possibly much lighter and, at the leading order, the heavier species $\chi/\bar{\chi}$ do not couple directly to the SM, resembles the case of  \textsl{assisted freeze-out} mechanism~\cite{Belanger:2011ww}. The relevant Boltzmann equations read then
\begin{align}
\label{eq:Bol1}
\frac{dn_\phi}{dt} + 3\,H\,n_\phi & = -\langle\sigma v\rangle_{\phi\phi\rightarrow \textrm{SM\,SM}}\left[n_\phi^2-(n_\phi^{\textrm{eq}})^2\right]\\
& \hspace{0.4cm}- \langle\sigma v\rangle_{\phi\phi\rightarrow\chi\bar{\chi}}\left[n_\phi^2-\left(\frac{n_\phi^\textrm{eq}}{n_\chi^{\textrm{eq}}}\right)^2 n_\chi^2\right],\nonumber
\\
\frac{dn_\chi}{dt} + 3\,H\,n_\chi & =  -\frac{1}{2}\langle\sigma v\rangle_{\chi\bar{\chi}\rightarrow\phi\phi}\left[n_\chi^2-\left(\frac{n_\chi^\textrm{eq}}{n_\phi^{\textrm{eq}}}\right)^2 n_\phi^2\right],
\label{eq:Bol2}
\end{align}
where we have simplified notations by using $n_\chi\equiv n_{\chi+\bar{\chi}}$. The first term on the RHS of \cref{eq:Bol1} corresponds to  the $\phi\phi$ annihilations into the SM particles, and the remaining terms are relevant for annihilations between the dark species. We leave a more detailed discussion of how to solve these equations to \cref{app:Boltzmann}, while here we briefly describe the most important features of the solutions, before discussing the results.

As can be seen from \cref{eq:Bol1}, as long as the $\chi$ particles are in thermal  equilibrium, $n_{\chi}\approx n_{\chi}^{\textrm{eq}}$, the $\phi$ number density undergoes a standard evolution with an additional impact from possible annihilations into $\chi$ DM that effectively add to the total thermally averaged annihilation cross section. In particular, as characteristic for any species that decouple whilst being relativistic, the yield of ultra-light $\phi$ particles in the conformal and disformal models occurs to be to roughly independent of the temperature, $Y_\phi=n_\phi/s \simeq (g_\phi/g_{\ast s})\,(45/2\pi^4)$, where $s$ and $g_{\ast s}$ are the entropy density and the relevant number of relativistic degrees of freedom, while $g_\phi=1$ is the number of degrees of freedom for $\phi$. For $\phi$ contributing to HDM, this leads to the following relic abundance of $\phi$
\begin{equation}
(\Omega_\phi h^2)_{\textrm{HDM}} \sim 10^{-3}\,\left(\frac{100}{g_{\ast s}(x_{\phi,\textrm{fo}})}\right)\,\left(\frac{m_\phi}{\textrm{eV}}\right),
\label{eq:Oh2relativistic}
\end{equation}
where $x_\phi = m_\phi/T_{\phi,\textrm{fo}}$. Instead, for a nearly massless scalar field, the relevant present-day contribution to the energy density of the Universe is governed by the model-dependent $\phi$ potential and related slow-roll or oscillations of the field. 

We note that a precise determination of $T_{\phi,\textrm{fo}}$ for ultra-light scalars would require going to a relativistic version of \cref{eq:Bol1} with the  Bose-Einstein statistics taken into account. In addition, one should also include effects from a possible non-zero thermal mass of $\phi$ that is induced by scalar self-couplings (see e.g. Refs~\cite{Olechowski:2018xxg,Arcadi:2019oxh} for recent similar discussions and \cref{app:Boltzmann} for further relevant comments). However, in the analysis below, we primarily focus on the relic density of heavy WIMPs $\chi$ that freeze-out being non-relativistic. As long as $n_\phi\approx n_{\phi}^{\textrm{eq}}$ close to the $\chi$ freeze-out, the relevant relic abundance is then dictated by a WIMP-like evolution employing the Maxwell-Boltzmann approximation, cf. \cref{eq:Bol2}, and is given by 
\begin{equation}
\Omega_\chi h^2\sim (0.1)\,\sqrt{\frac{100}{g_{\ast}(x_{\chi,\textrm{fo}})}}\,
\left(\frac{x_{\chi,\textrm{fo}}}{20}\right)\,\left(\frac{10^{-9}\,\textrm{GeV}^{-2}}{\langle\sigma v\rangle}\right),
\end{equation}
where $x_{\chi,\textrm{fo}}=m_\chi/T_{\chi,\textrm{fo}}$,  while $g_{\ast}(x_{\chi,\textrm{fo}})$ is the number of relativistic degrees of freedom at $T_{\chi,\textrm{fo}}$. 

\subsection{Results for conformal and disformal models\label{sec:results}}

\begin{figure}[tb]
\includegraphics[width=0.4\textwidth]{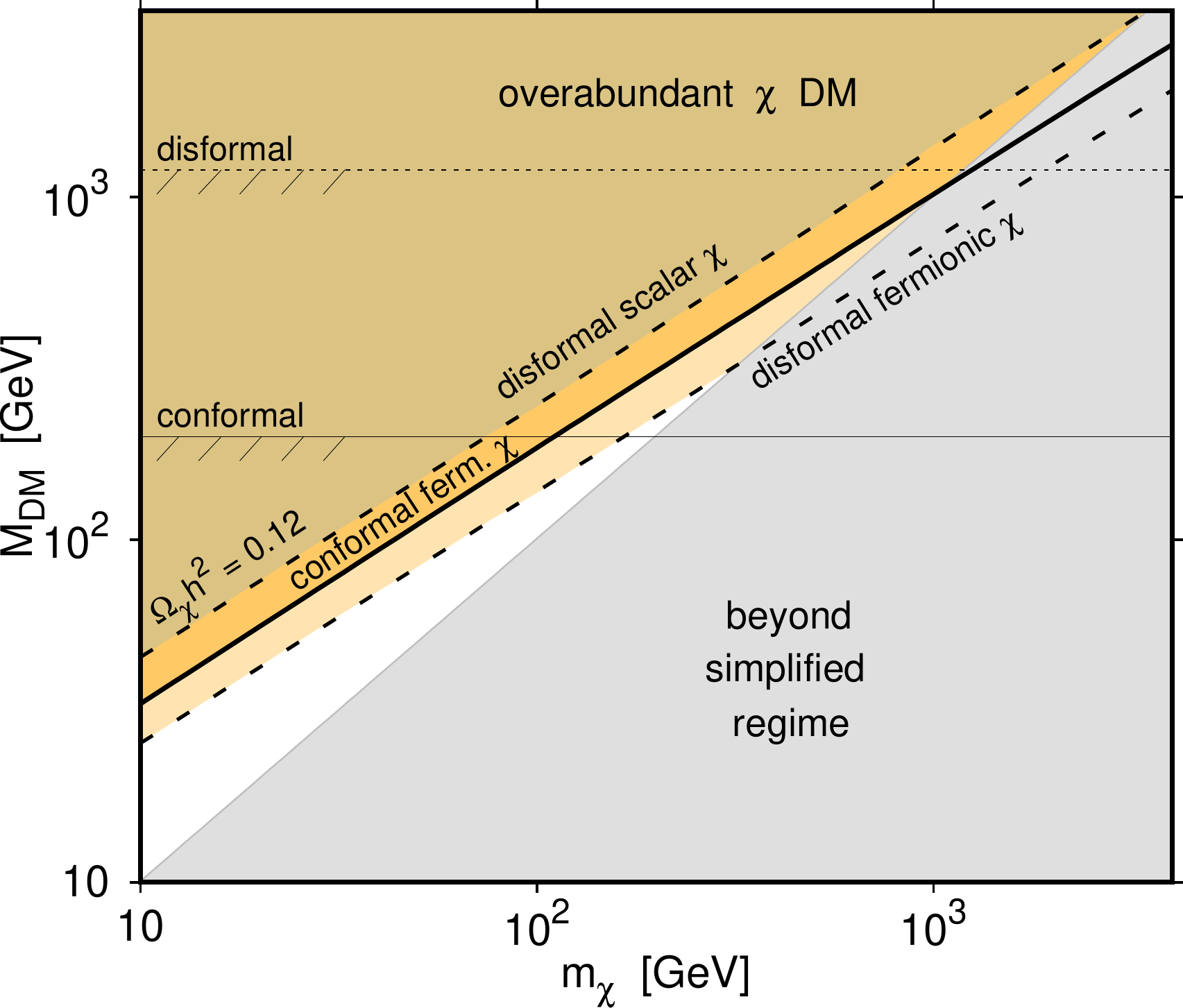}
\caption{The lines in the $(m_\chi,M_{\textrm{DM}})$ parameter space of the conformal (solid) and disformal (dashed) models that yield the $\chi$ relic density which is equal to the observed abundance of DM. The results are shown for fermionic $\chi$ particles. In addition, the scalar $\chi$ DM case is shown with the upper dashed line for the disformal model. The brown- and orange-shaded regions in the plot that lie above each of the lines, respectively, correspond to a large thermal relic over-abundance of $\chi$, while the white region below the lines predicts $\Omega_\chi h^2$ below the Planck limits. The gray-shaded region indicates a regime in which the EFT approach ceases to be valid, which is roughly characterized by $m_\chi\gtrsim M_{\textrm{DM}}$. Horizontal solid (dotted) line corresponds to the LHC limits on $M_{\textrm{DM}} = M_{\textrm{SM}}$ for conformal (disformal) coupling.}
\label{fig:MDMmchi}
\end{figure}

In \cref{fig:MDMmchi}, we show the lines of constant $\chi$ DM relic abundance, $\Omega_\chi h^2\simeq 0.12$~\cite{Aghanim:2018eyx}, obtained by solving the Boltzmann \cref{eq:Bol1,eq:Bol2}. From top to bottom, the lines correspond to disformally coupled scalar DM, conformally coupled Dirac fermion DM, and disformally coupled fermionic $\chi$, as dictated by the relevant strength of their couplings. In each case, the correct $\chi$ DM relic density can be obtained for $m_\chi^6/M_{\textrm{DM}}^8\sim (10^{-6}-10^{-5})~\textrm{GeV}^{-2}$, cf. discussion in \cref{sec:interactions}. 

As the DM mass grows, $m_\chi\gtrsim \textrm{several TeV}$, the ratio $m_\chi^6/M_{\textrm{DM}}^8$ becomes too small unless we allow $M_{\textrm{DM}}\lesssim m_\chi$, which violates the approximate validity conditions of the EFT approach given in \cref{eq:MDMlimit}. This is indicated by a gray-shaded region in the plot. However, both these conditions and the $\chi$ relic abundance constraint can be simultaneously satisfied for a range of lower DM masses, $m_\chi\lesssim 3.8\,\textrm{TeV}, 1\,\textrm{TeV}, 300\,\textrm{GeV}$, respectively, for the three aforementioned models.

Notably, the precise value of the ultra-light scalar mass $m_\phi$, as well as the parameter $M_{\textrm{SM}}$ which governs its coupling strength to the SM, do not play important roles in obtaining the $\chi$ DM relic abundance. In particular, $M_{\textrm{SM}}$ should be sufficiently small so that $\phi$ remains in thermal equilibrium with the SM around the freeze-out of $\chi$. This, however, is guaranteed for the range of values of this parameter above the lower limits given by \cref{eq:MSMLHCconf,eq:MSMLHCdisf}. As discussed in \cref{sec:thersigmavTfo}, the decoupling temperature of the $\phi-\textrm{SM}$ interactions grows almost linearly with the increasing $M_{\textrm{SM}}$ parameter, $T_{\phi,\textrm{SM}}\sim M_{\textrm{SM}}^{8/7}/M_{\textrm{Pl}}^{1/7}$. In particular, assuming $M_{\textrm{SM}}\simeq M_{\textrm{DM}}$ along the lines with the correct $\chi$ relic density, the value of $T_{\phi,\textrm{fo}}$ does not exceed the $\chi$ DM freeze-out temperature for $m_\chi\gtrsim 100~\textrm{GeV}$ in \cref{fig:MDMmchi}.

By taking into account the lower bounds on $M_{\textrm{SM}}$ from the LHC, as shown with horizontal lines in \cref{fig:MDMmchi}, and the upper bound $m_\chi\lesssim M_{\textrm{DM}}$ dictated by the EFT validity, one can obtain the correct value of $\chi$ relic density assuming universal coupling, $M_{\textrm{SM}}=M_{\textrm{DM}}$, for the following mass ranges:
\begin{align}
\label{eq:massrangeconf}
100~\textrm{GeV} &\lesssim m_{\chi}\lesssim 1~\textrm{TeV}\hspace{0.18cm}\textrm{(conformal, fermionic $\chi$)},\\
800~\textrm{GeV} &\lesssim m_{\chi}\lesssim 3.8~\textrm{TeV}\hspace{0.18cm}\textrm{(disformal, scalar $\chi$)}. 
\label{eq:massrangedisf}
\end{align}
In this case, as well as for a more general scenario with $M_{\textrm{DM}}$ lying below the lines with $\Omega_\chi h^2\simeq 0.12$ in \cref{fig:MDMmchi}, ultra-light scalars conformally or disformally coupled to matter would have an inevitable impact on the WIMP DM relic density. In particular, the $\phi$ portal can solely provide the correct DM relic density, even if other couplings of $\chi$ to the SM are suppressed. It can also contribute to $\Omega_\chi h^2$  non-negligibly  in scenarios in which the other couplings are tuned to provide the DM relic abundance close to the value reported by the Planck collaboration~\cite{Aghanim:2018eyx}. 

Instead, if $M_{\textrm{DM}}$ grows to larger values, the resulting $\chi$ DM relic density obtained due to the interactions with $\phi$ remains too large, as indicated by brown and orange-shaded regions in \cref{fig:MDMmchi}. The DM abundance could then be driven to the observed value thanks to other couplings of $\chi$, with a negligible role played by  the $\phi$ portal. This also remains the case for the scenario with $M_{\textrm{DM}}= M_{\textrm{SM}}\gtrsim 1.2~\textrm{TeV}$ and disformally coupled fermionic $\chi$, where a substantial impact of the interactions with $\phi$ on the WIMP $\chi$ relic abundance would require assuming non-universal couplings with $M_{\textrm{DM}}$ about an order of magnitude lower than $M_{\textrm{SM}}$. We note, however, that a more detailed treatment of the EFT validity limit and a possible UV completion could extend the allowed DM mass regime. A significant impact on WIMP DM relic density could then be obtained for the universal coupling to matter. This would correspond to a mass of disformally coupled fermionic DM above $2~\textrm{TeV}$. 

\section{Phenomenology of derivative conformal and disformal couplings\label{sec:pheno}}

Given the possible important impact of conformally and disformally coupled ultra-light scalars on the DM relic density, it is relevant to discuss the phenomenological implications of such a scenario.
\paragraph{$\phi$-SM interactions} As discussed in \cref{sec:bounds}, the most stringent constraints on the conformal and disformal energy scales $M_{\textrm{SM}}$ come from the null LHC searches for light scalars~\cite{Brax:2016did}, with the dominant bounds recently published by the ATLAS collaboration~\cite{Aaboud:2019yqu}. These are based on $36.1~\textrm{fb}^{-1}$ of data collected during an initial part of the LHC Run 2. In total, one expects to collect up to $\sim 10$ times more data after Run 3, and even up to a factor of $\sim 100$ more data once a future High Luminosity LHC (HL-LHC) phase is taken into account~\cite{ApollinariG.:2017ojx}. Given a strong dependence of the production cross section on the conformal and disformal energy scales, $\sigma\sim1/ M^{8}_{\textrm{SM}}$, this translates into an expected improvement in the relevant bounds on $M_{\textrm{SM}}$, cf. \cref{eq:MSMLHCconf,eq:MSMLHCdisf}, by a factor $\lesssim 2$.

Assuming universal couplings of $\phi$ to matter, $M_{\textrm{SM}}=M_{\textrm{DM}}$, the future LHC sensitivity to the $M_{\textrm{SM}}$ parameter could be interpreted in terms of the coupling to DM. In particular, in the full run of HL-LHC, one will be able to probe a large fraction of the mass range relevant for disformally coupled scalar DM with $\Omega_\chi h^2\simeq 0.12$, cf. \cref{eq:massrangedisf}, as well as a part of the region corresponding to conformally coupled fermionic DM characterized by $m_\chi\lesssim 300~\textrm{GeV}$, cf. \cref{eq:massrangeconf}. 

In the presence of other couplings of $\chi$ to the SM, searches for $\phi$ missing energy signature could be further complemented by similar studies for $\chi$, depending on the model. In addition, further bounds or hints of the existence of new light scalars could be deduced from the SM precision physics, e.g. the $(g-2)_\mu$ anomaly~\cite{Cembranos:2005jc,Cembranos:2005sr}.

As discussed in \cref{sec:bounds}, light scalars could also contribute to the total energy density of relativistic species in the early Universe, with a typical impact on $N_{\textrm{eff}}$ corresponding to \cref{eq:DelNeff} with $\Delta N_{\textrm{eff}}\lesssim 0.06$. Interestingly,  such a low value of $\Delta N_{\textrm{eff}}$ could manifest itself in future CMB surveys, although generally at the level of only about $1\sigma$ deviation from the standard cosmological model~\cite{Abazajian:2016yjj}. Such an  excess in $N_{\textrm{eff}}$ happens for the entire range of the $M_{\textrm{SM}}\simeq M_{\textrm{DM}}$ parameter that is relevant to our discussion in \cref{sec:results}.

\paragraph{$\phi$-DM interactions} On the other hand, searches for direct and indirect signatures of WIMP DM lie outside the high-energy and precision frontiers and, therefore, often remain unaffected by the presence of derivative conformal and disformal couplings of light scalars. In particular, as briefly discussed in \cref{app:otherinteractions,app:SE}, both $\phi$-loop-induced couplings and the Sommerfeld enhancement do not lead to an observable increase in the $\chi$ DM interaction rates, unless one probes regions of the parameter space of the models that lie close to, or possibly beyond, the validity regime of the EFT approach. Hence, if the $\chi$s are coupled to the SM solely via the light $\phi$, then they remain secluded, as both their annihilation rates into visible final states and scattering cross sections off SM species are much suppressed. 

On the other hand, in more general WIMP scenarios, the low-energy phenomenology can be driven by other couplings between the SM and $\chi$ DM, if they generate detectable signatures. This detection prospects, however, can also be affected by the presence of the $\chi$ DM couplings to light $\phi$s. In particular, the disformally coupled scalar DM particles have unsuppressed $s$-wave annihilation channels into light scalars, cf.~\cref{eq:sigmaannswave}. As a result, a fraction of such $\chi$s will decay invisibly, therefore reducing the expected signal rates in indirect DM searches, while the $\phi$ impact on direct detection of $\chi$ DM remains negligible. Notably, the annihilation process into $\phi\phi$ pairs can create a flux of boosted $\phi$s produced e.g. at the GC. This, however, for $m_\chi\sim \textrm{TeV}$, remains essentially undetectable in direct searches due to tiny scattering cross sections, cf.~\cref{app:scattering}.

We leave a more detailed analysis of the interplay between $\phi$-induced couplings and other portals to DM in specific WIMP models for future studies.

\paragraph{Propagation of gravitational waves} An additional phenomenological aspect of the models considered in this paper appears when they are viewed as theories of modified gravity. This can be seen by writing the gravitational sector, which we assumed to be the Einstein--Hilbert action plus a scalar field $\phi$, in terms of the standard model metric. In this frame, the theory becomes a Horndeski theory, cf.~\cref{app:EMgravwaves}. As a consequence, in general, gravitational waves might not propagate with the speed of light $c$~\cite{Kobayashi:2011nu}. This can lead to detectable signatures in multi-messenger astronomy that employs both gravitational-wave searches and other types of observations. 

A pure conformal coupling does not generate any deviation of the gravitational wave speed from $c$, but a disformal coupling can have a larger impact.
In particular, in the case of a pure disformal coupling to the standard model sector with $D_{\rm SM} = 2/M_{\rm SM}^4$ in \cref{eq:metric}, the speed of gravitational waves is given by (see~\cref{app:EMgravwaves})
\begin{equation}
c_T^2 = 1 - 2\,\dot\phi^2/M_{\rm SM}^4.
\label{eq:cT}
\end{equation}
On the other hand, the recent observations of the neutron star merger GW170817~\cite{TheLIGOScientific:2017qsa} and its electromagnetic counterpart GRB170817A~\cite{Goldstein:2017mmi,Savchenko:2017ffs} constrain $|c_T-1|$ to be smaller than $10^{-15}$. This can then be rewritten as a bound on the $M_{\textrm{SM}}$ parameter.

Let us first assume that $\phi$ evolves slowly today with $\dot \phi = \alpha H_0 M_{\rm Pl}$, where $\alpha={\cal O}$(1), $H_0 \approx 10^{-42}$~GeV and $M_{\rm Pl}$ is the reduced Planck mass. By substituting this to \cref{eq:cT}, we find a current bound $M_{\rm SM}>10^{-8}$~GeV. Notably, this is a very weak constraint in comparison with the aforementioned LHC bounds and expected future reach. If, instead, the scalar field oscillates quickly around the minimum of the potential, it contributes to the DM energy density. For a harmonic potential with $V = m_\phi^2 \phi^2/2$, we have $\rho_{\rm kin}\sim \rho_{\rm pot}$ and hence we can estimate $\rho_\phi\approx\rho_{\rm kin}=\langle \dot\phi^2 \rangle$. If the density of the $\phi$-particles is of the order of $\rho_{\rm CDM}$, then at the present time we have $\langle \dot\phi^2 \rangle_0\sim \rho_{\rm cr,0}\,\Omega_{\rm CDM}\approx 10^{-47}~{\rm GeV}^4$. Again, by employing \cref{eq:cT}, one obtains $M_{\rm SM}>10^{-8}$~GeV as it is the case for a slowly-rolling field. This is not surprising, since in both cases the time-derivative of $\phi/M_{\rm Pl}$ is of order $H_0$. If the contribution of the scalar field to dark matter is negligible, the constraint will be even weaker. 

Last, but not least, we stress that the impact of light $\phi$ remains also beyond capabilities of future multi-messenger searches, since for them to become competitive with the LHC, one would require several tens of orders of magnitude improvement in the bound on the gravitational-wave speed, $|c_T-1|\lesssim 10^{-55}$.

\section{Conclusions\label{sec:conclusions}}

Light new scalar $\phi$ degrees of freedom are prevalent in cosmology, as well as in various BSM models with possibly profound consequences for astrophysics. In particular, they have recently gained renewed attention in connection to the swampland conjecture~\cite{Obied:2018sgi} and the nature of the dark energy content of the Universe, while they appear also in many other contexts. If such fields exist, they could also naturally be coupled to matter.

A particularly interesting example of such a scenario employs shift-symmetry-preserving derivative couplings of $\phi$. If such symmetry is only very softly broken for $m_\phi\neq 0$, then one expects that the dominant couplings of $\phi$ to other particles are induced by the aforementioned derivative operators.

In this study, we have discussed the phenomenology of such conformal and disformal couplings that are assumed to be nearly or strictly universal to all matter species. These interactions can naturally emerge from scenarios of modified gravity, in which both the SM and DM particles move along geodesics altered by $\phi$-induced couplings. At a more general level, they can be considered as an example of higher-dimensional operators providing a portal to DM and allowing simple discrete symmetries to stabilize it.

The derivative couplings under study cause a strong energy-dependence of the interaction cross sections. As a result, the couplings effectively switch on in collider searches or at large temperatures characteristic for the early Universe, while typically remain screened from low-energy tests. In particular, we have shown that taking into account the current bounds from the LHC and the validity regime of a simple EFT approach, the interaction rates can be strong enough to remain in thermal equilibrium around typical WIMP freeze-out. If present, they would then have an inevitable impact on $\chi$ WIMP DM relic density, with a special emphasis on scenarios with $100~\textrm{GeV}\lesssim m_\chi\lesssim (\textrm{a few})~\textrm{TeV}$. Importantly, such scenarios could be independently tested in collider missing-energy searches for $\phi$, as well as they can leave a mild impact on the $N_{\textrm{eff}}$ parameter measured in future CMB surves.

We have shown that  such a portal to the SM could lead to the correct DM relic density of fermionic or scalar $\chi$ on its own. Importantly, if other types of interactions between DM and the SM are also effective in the early Universe, light scalar conformal and disformal portals could still give important contributions to the total annihilation rate.  The impact of $\phi$ on the present-day phenomenology of $\chi$ DM could also be non-negligible. In particular, scalar DM particles can have an efficient annihilation channel into the invisible $\phi\phi$ pair, which is not suppressed for small velocities. On the other hand, DM direct detection prospects would be driven by other non-$\phi$-induced interactions. 

Although light conformally or disformally coupled scalars can be well-motivated independently of models with heavy WIMP DM, both types of species can appear simultaneously in more complex scenarios. In particular, in brane-world models, disformally coupled light scalars can naturally emerge on top of heavier such species playing the role of CDM (see~\cite{Cembranos:2016jun} for a recent review). On the other hand, ultra-light oscillating scalar fields with $m_\phi\sim 10^{-22}~\textrm{eV}$ could also behave like CDM themselves, as for fuzzy or scalar field DM (SFDM), see Refs.~\cite{Hui:2016ltb,Lee:2017qve,Ferreira:2020fam} for recent reviews. If such fields are coupled to matter, as discussed in our analysis, they can be active at an early epoch of the evolution of the Universe, even before rapid CDM-like oscillations begin. This could lead to a mixed scenario with both fuzzy and WIMP DM, with the latter being coupled to the SM even only through a `fuzzy' conformal or disformal portal. We leave a detailed analysis of this scenario for future studies.

The existence of ultra-light scalar fields is particularly well-motivated in cosmology. Similarly, as far as models of DM are concerned, the dominant paradigm is to search for new heavy WIMP-like species. A possible interplay between these two extensions of the SM can have fundamental cosmological and astrophysical consequences. A more general analysis and a better connection to fundamental physics which would involve both ingredients is certainly welcome and deserves further study.

\acknowledgements We would like to thank Jose Cembranos and Andrzej Hryczuk for helpful comments on the manuscript. ST would like to thank Jonathan Feng for useful discussions. ST would like to express special thanks to the Mainz Institute for Theoretical Physics (MITP) of the Cluster of Excellence PRISMA+ (Project ID 39083149) and the GGI  Institute  for  Theoretical  Physics  for  their  hospitality  and  support.  CvdB and ST are supported by  the  Lancaster-Manchester-Sheffield  Consortium  for Fundamental Physics under STFC grant ST/P000800/1. ST is partially supported by the Polish Ministry of Science and Higher Education through its scholarship for young and outstanding scientists (decision no. 1190/E-78/STYP/14/2019). This article is based upon work related tothe COST Action CA15117 (CANTATA) supported byCOST (European Cooperation in Science and Technology).

\appendix

\section{Loop-induced interactions and scalar self-couplings\label{app:otherinteractions}}

In this appendix, we briefly discuss additional interactions that can be present in the models under study, which, however, play a subdominant role in setting the relic density of WIMP-like $\chi$ DM.

\paragraph{Loop-induced couplings} Besides direct interactions with the scalar field $\phi$, the $\chi$ DM and SM particles also have loop-induced couplings to each other. However, the resulting cross sections are typically suppressed and, therefore, play a negligible role in our discussion.

In order to illustrate this, let us focus on the case of the disformal coupling that has been studied in Ref.~\cite{Cembranos:2005jc}. The loop-induced effective four-point interaction between fermions in this model is given by
\begin{equation}
\mathcal{L} = \frac{\Lambda^4}{3072\,\pi^2\,M^8}\,\left(2\,T_{\mu\nu}\,T^{\mu\nu}+T_\mu^\mu\,T_\nu^\nu\right),
\label{eq:loopdisf}
\end{equation}
where the explicit cut-off scale $\Lambda$ has been introduced to deal with a non-renormalizable nature of the disformal coupling. One requires $\Lambda<4\sqrt{\pi}\,M$ for a perturbative treatment of the loop expansion. Assuming a universal suppression scale $M_{\textrm{SM}}= M_{\textrm{DM}}\equiv M$, one obtains from \cref{eq:loopdisf} the cross section of direct $\chi$ annihilation into SM fermions, $\chi\bar{\chi}\rightarrow f\bar{f}$, of order $\sigma_{\textrm{dir.ann.}}v\sim 10^{-10}\,\Lambda^8\,m_\chi^6\,v^2 / M^{16}$. For $\Lambda\sim m_\chi\sim M\sim \textrm{TeV}$, this yields the value of $\sigma_{\textrm{dir.ann.}}$ several orders of magnitude lower than the typical thermal cross section. We therefore neglect such direct annihilation processes hereafter, when discussing $\chi$ DM production in the early Universe. 

The suppression is even stronger when DM annihilation in the GC are considered at present times, which is due to small DM velocities, $v\sim 10^{-3}$. Notably, the Sommerfeld enhancement induced by a long-range force between DM particles mediated by $\phi$, is typically mild, $S\sim 1$, cf.~\cref{app:SE}. Last but not least, a loop-induced $\chi$ scattering off SM fermions, $\chi f\rightarrow \chi f$, also remains below a measurable level, unless one focuses on regions of the parameter space with $\Lambda>M$ that probe scenarios where the model-dependent impact of possible UV completions should begin to play a non-negligible role in the analysis.

A large excess of $\Lambda$ over $M$ can additionally be constrained by precision studies. A notable example is the corresponding impact on the muon anomalous magnetic moment, $(g-2)_\mu$, which grows with increasing ratio $\Lambda/M$. On the other hand, for $M^4/\Lambda^3\sim \mathcal{O}(1-10~\textrm{GeV})$, the disformal-loop contributions might accommodate for the entire observed experimental excess~\cite{Cembranos:2005jc,Cembranos:2005sr}. The model would then be subject to bounds in the light of new expected data. 

\paragraph{Scalar self interactions} Scalar self interactions can also be induced at a loop level via virtual SM or DM particles. On top of this, if the scalar field couplings to matter are induced from an action containing $\phi$-dependent metric, cf. \cref{eq:metric}, additional scalar self-interaction terms appear once the respective energy-momentum tensor for $\phi$ is taken into account in \cref{eq:Lphiconf,eq:Lphidisf}. Similarly to the couplings of $\phi$ to the SM and DM, the dominant  terms are suppressed by $1/M^4$ and contain terms that couple four scalar fields, e.g. $(\partial\phi)^2(\partial\phi)^2$ or $(\partial\phi)^2\phi^2$~\cite{BeltranJimenez:2018tfy}. Such interactions contribute to the effective momentum transfer in the $\phi$ sector in the early Universe. On the other hand, their impact on the $\phi$ number density via possible $2\leftrightarrow 4$ processes occurs only at the next-to-leading order and is suppressed with respect to the interactions with the SM particles. We then neglect such processes below, when analyzing the relevant Boltzmann equations.

\section{Details about the Boltzmann equations\label{app:Boltzmann}}

\subsection{Thermally averaged cross section\label{sec:thersigmavTfo}}

When solving the Boltzmann equations, an important quantity that determines both the freeze-out temperature and relic abundance is the thermally averaged cross section. In case of an approximate Maxwell-Boltzmann statistics, it can be written as~\cite{Gondolo:1990dk}
\begin{align}
\label{eq:theraver}
\langle\sigma_{\textrm{ann}} v\rangle_{T} =& \frac{1}{8\,T\,\left[m_\chi^2\,K_2\left(\frac{m_\chi}{T}\right)\right]^{2}} \\
&\hspace{-1.5cm}\times \int_{\max{\{4m_\chi^2,4m_f^2}\}}^{\infty}{ds\,\left[\sigma_{\textrm{ann}}\times(s-4m_\chi^2)\,\sqrt{s}\,K_1\Big(\frac{\sqrt{s}}{T}\Big)\right]},\nonumber
\end{align}
where $K_1$ and $K_2$ are modified Bessel functions of the second kind. In the relativistic limit, the proper quantum statistics should be used that can modify the collision rate.

The decoupling temperature of efficient annihilations of $\phi$ can be estimated by comparing the annihilation rate $n^{\textrm{eq}}_\phi\,\langle\sigma v\rangle\sim T_{\textrm{fo}}^9/M^8$, which should be summed over all the relevant SM degrees of freedom, to the expansion rate of the Universe, $H(T_{\textrm{fo}})\simeq (1.67\sqrt{g_{\textrm{eff}}})\,T_{\textrm{fo}}^2/M_{\textrm{Pl}}$. In case of relativistic scalar particles $\phi$ with a negligible mass, the typical  temperatures for $M_{\textrm{SM}}$ satisfying bounds from \cref{eq:MSMLHCconf,eq:MSMLHCdisf}, are about GeV and slowly grow with increasing value of this parameter, $T_{\phi,\textrm{fo}}\sim M_{\textrm{SM}}^{8/7}/M_{\textrm{Pl}}^{1/7}$. We have checked that this is only mildly affected when proper Bose-Einstein statistics is taken into account for $\phi$ in the annihilation cross section, cf. Ref.~\cite{Olechowski:2018xxg}.

At the time of $\chi$ decoupling, the dominant contributions to $\langle\sigma v\rangle$ for conformally and disformally coupled scalars $\phi$ annihilating into the SM species are the following:
\begin{itemize}
\item \textsl{Conformal model} Due to the proportionality to the mass of the fermion in the annihilation cross section, cf. \cref{eq:sigannconf}, the value of $\langle\sigma v\rangle$ is driven by the annihilations to the heaviest kinematically available fermions, for a given center-of-mass energy $\sqrt{s}$. The relevant expressions read
\begin{equation}
\sigma_{\textrm{ann}}(\phi\phi\rightarrow f\bar{f}) = 8\,\left(1-\frac{4m_f^2}{s}\right)\times \sigma_{\textrm{ann}}(f\bar{f}\rightarrow\phi\phi),
\end{equation}
where the annihilation cross section of the inverse process, $f\bar{f}\rightarrow\phi\phi$, is given in \cref{eq:sigannconf}. 

\item \textsl{Disformal model} The contributions from the annihilations into all kinematically available SM fermions play important role, as well as the one related to gluons in the final state
\begin{equation}
\sigma_{\textrm{ann}}(\phi\phi\rightarrow gg) = \frac{s^3}{240~\pi\,M_{\textrm{SM}}^8}.
\label{eq:gluons}
\end{equation} 
\end{itemize}

As far as $\chi$ DM annihilations into a $\phi\phi$ pair are concerned, the relevant annihilation cross sections are given in \cref{eq:sigannconf,eq:siganndisf,eq:siganndisfscalar}. Importantly, for $m_\chi\gtrsim 100~\textrm{GeV}$, at the time of $\chi$ freeze-out, a $\phi$-mediated kinetic equilibrium between the SM and $\chi$ DM sectors is maintained. We, therefore, neglect the effects of a possible early kinetic decoupling when solving the Boltzmann equations. 

\subsection{Solving the Boltzmann equations}

It is useful to rewrite the Boltzmann equations \cref{eq:Bol1,eq:Bol2} in terms of the yields of dark species, $Y_{\chi/\phi}=n_{\chi/\phi}/s$, and the $x_\phi=m_\phi/T$ variable~\cite{Belanger:2011ww}
\begin{align}
\frac{dY_\phi}{dx_\phi} &= -\frac{\lambda_{\phi\phi\rightarrow \textrm{SM\,SM}}}{x_\phi^2}\left[Y_\phi^2-(Y_\phi^{\textrm{eq}})^2\right]\nonumber\\
&\hspace{0.38cm}+\frac{\lambda_{\chi\bar{\chi}\rightarrow\phi\phi}}{x_\phi^2}\left[Y_\chi^2-\left(\frac{Y_\chi^\textrm{eq}}{Y_\phi^{\textrm{eq}}}\right)^2 Y_\phi^2\right],
\label{eq:Bol1Y}
\\
\frac{dY_\chi}{dx_\phi} & =  -\frac{\lambda_{\chi\bar{\chi}\rightarrow\phi\phi}}{x_\phi^2}\left[Y_\chi^2-\left(\frac{Y_\chi^\textrm{eq}}{Y_\phi^{\textrm{eq}}}\right)^2 Y_\phi^2\right],
\label{eq:Bol2Y}
\end{align}
where we have used the fact that the second term on the RHS of \cref{eq:Bol1Y} has to compensate for the RHS of \cref{eq:Bol2Y}, and we have introduced
\begin{equation}
\lambda = \left(\frac{x s}{H}\right)\,\left(\frac{1}{2}\langle\sigma v \rangle_{\chi\bar{\chi}\rightarrow\phi\phi}\right).
\label{eq:lambda}
\end{equation}
In \cref{eq:lambda}, for the radiation dominated epoch, $s(T) = g_{\ast\,s}\,(2\pi^2/45)\,T^3$ and $H(T)=\sqrt{\rho(T)/(3\,M_P^2)}$ with $\rho\simeq\rho_R = g_{\ast}(T)\,(\pi^2/30)\,T^4$. 

For each of the dark species, the equilibrium yield is given by
\begin{equation}
Y_{\phi/\chi}^{\textrm{eq}} = \frac{g_{\phi/\chi}}{g_{\ast\,s}}\,\frac{45}{4\pi^4}\,(x_{\phi/\chi})^2\,K_2(x_{\phi/\chi}).
\end{equation}
In particular, for a very small mass of $\phi$ we obtain
\begin{equation}
Y_\phi^{\textrm{eq}} \stackrel{m_\phi\rightarrow 0}{\rightarrow} \frac{g_\phi}{g_{\ast\,s}}\,\frac{45}{2\pi^4},
\label{Yphilight}
\end{equation}
i.e. the equilibrium yield remains roughly independent of $T$ (beside a mild temperature dependence of $g_{\ast\,s}$). As a result, for a negligible mass of $\phi$, to a good approximation $Y^{\textrm{eq}}_\phi\approx Y_\phi$ in \cref{eq:Bol2Y}, and the evolution of the $\chi$ DM relic density resembles the one of a standard WIMP-like scenario with the annihilation cross section dictated by the process $\chi\bar{\chi}\rightarrow\phi\phi$.

\section{Sommerfeld enhancement for conformally and disformally coupled dark matter\label{app:SE}}

Light intermediate bosonic particles mediating DM self-interactions can lead to a substantial increase of the annihilation and self-scattering cross sections once they induce long-range attractive forces. This effect, known as the Sommerfeld enhancement (SE) of DM interactions~\cite{Hisano:2004ds}, becomes the most prominent in the non-relativistic limit, in which quasi-bound states can be formed.

\begin{figure}[t]
\includegraphics[width=0.4\textwidth]{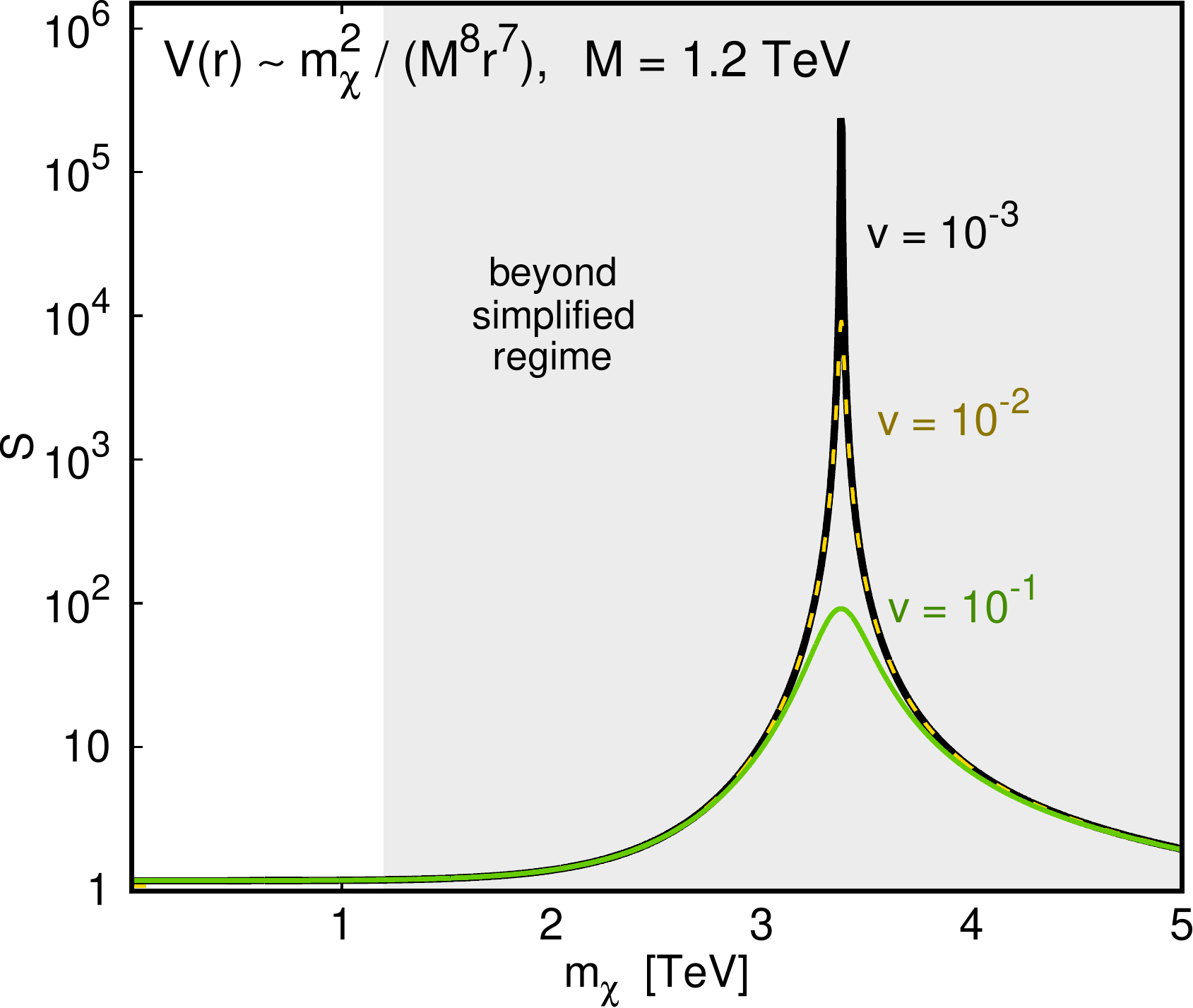}
\caption{Sommerfeld enhancement factor $S$ for the potential corresponding to a disformal coupling between two DM particles, as a function of the DM mass $m_\chi$. The disformal mass scale is equal to, $M=1200~\textrm{GeV}$. The results are shown for three values of the velocity, $v = 10^{-1}, 10^{-2},10^{-3}$ with solid black, dashed golden and solid green lines, respectively. For velocities smaller than $v\sim 10^{-3}$, the results are similar to the black solid line. The relevant results for other values of $M$ and for conformal couplings are similar, but shifted in $m_\chi$ such that the first peak corresponds to $m_\chi\sim 3\,M$.}
\label{fig:SErminus7}
\end{figure}

The relevant  force between a pair of DM particles $\chi$ coupled through a $\phi$-portal corresponds to the following potential~\cite{Kaloper:2003yf}
\begin{equation}
V(r) = -\frac{3\,A}{32\,\pi^3\,r^7}\,\frac{m_\chi^2}{M^8},
\label{eq:Vdisffermions}
\end{equation}
where $A=20$ for conformally coupled fermions~\cite{Brax:2017xho} and $A=1$ ($5/2$) for disformally coupled fermions~\cite{Brax:2014vva} (complex scalars), respectively. We denote by $M=M_{\textrm{DM}}=M_{\textrm{SM}}\gtrsim 200~\textrm{GeV}$ ($1200~\textrm{GeV}$) the (conformal) disformal mass scale. 

We note that, for typical DM velocities in dwarf galaxies and the GC, $v\sim 10^{-5}-10^{-3}$, the potential becomes strongly coupled, $V(r)\sim m_\chi v^2/2$, only at very low distances, $r\sim (\textrm{a few})/M$, where we expect corrections going beyond the EFT approach to become important. In the following, we then assume, for simplicity, that a weakly coupled regime is valid until $r\gtrsim r_0 = 1/M$. For larger distances, the potential quickly vanishes with increasing $r$. It becomes effectively suppressed at a distance $r_\phi\sim\textrm{(a few)}\times r_0$, which is much smaller than the typical de Broglie wavelength of a WIMP DM particle, i.e. $r_\phi m_\chi v \ll 1$ for $m_\chi\lesssim M$. 

At the quantum level, the behavior of a system of two DM particles interacting with the potential $V(r)$ is determined by solving the relevant Schrodinger equation for the radial wavefunction. The SE factor can then be obtained by comparing the probability density of finding both particles at the same place, calculated with and without the potential, $S = \left|\psi/\psi_0\right|^2_{r\rightarrow 0}$. In \cref{fig:SErminus7}, we show such SE factor corresponding to the $\ell=0$ term in the partial wave expansion, as a function of the DM mass. When solving the relevant Schrodinger equation, we have regularized the singular potential $V\sim r^{-7}$ by replacing it for $r<r_0$ with a square well with height $V_0=V(r_0)$~\cite{Beane:2000wh}. In a more detailed treatment, it can be shown that by adjusting $V_0$ as a function of the cutoff distance and requiring a smooth transition of the wavefunction between the two regimes of the potential, $S$ can be kept independent of the choice of the cutoff~\cite{Bedaque:2009ri}. Last but not least, we have normalized the SE factor so that $S \rightarrow 1$ in the relativistic limit~\cite{Bellazzini:2013foa}.

As can be seen in \cref{fig:SErminus7}, for DM masses $m_\chi\lesssim M$, one obtains only a very mild enhancement, $S\simeq 1$, as expected for the quickly vanishing potential under study. The enhancement can grow for increasing DM mass, especially close to the resonance peaks. The first such peak corresponds to $m_\chi\sim 3\,M$. This, however, already lies in the regime in which the EFT description of the microscopic interactions  breaks down, and in which the potential becomes strongly coupled at larger distances, $r> r_0$. A detailed treatment of such a scenario goes beyond the scope of this study. On the other hand, for the region of the parameter space of  interest here, one can effectively assume that $S\simeq 1$.

\section{Scattering cross sections of conformally and disformally coupled scalars\label{app:scattering}}

The scattering cross section of a scalar $\phi$ with negligible mass off Dirac fermion $f$ with mass $m_f$ is given by
\begin{align}
\sigma_{\textrm{scat,C}} =& \frac{s^3}{64\,\pi\,M_C^8}\,\left(1-\frac{m_f^2}{s}\right)\label{eq:scatsigmafullconf}\\
&\times\left[2+7\frac{m_f^2}{s}-7\frac{m_f^4}{s^2}+25\,\frac{m_f^6}{s^3}-11\,\frac{m_f^8}{s^4}\right.\nonumber\\ 
&\hspace{0.5cm}+ \left.\left(6\,\frac{m_f^2}{s}+10\,\frac{m_f^4}{s^2}\right)\ln{\left(\frac{s}{m_f^2}\right)}\right],\nonumber
\end{align}
for the conformal model, cf. \cref{eq:Lphiconf}, and
\begin{align}
\sigma_{\textrm{scat,D}} =& \frac{s^3}{1536\,\pi\,M_D^8}\,\left(1-\frac{m_f^2}{s}\right)^4\label{eq:scatsigmafulldisf}\\
& \times\left[17+17\frac{m_f^2}{s}+11\,\frac{m_f^4}{s^2}+3\,\frac{m_f^6}{s^3}\right],\nonumber
\end{align}
for the disformal one, cf. \cref{eq:Lphidisf}. 

Both above expressions vanish in the non-relativistic limit with $x = E_\phi/m_f\ll 1$ (i.e. $s\simeq m_f^2$), where $E_\phi$ is the energy of the incident scalar in the rest frame of the initial-state fermion, although the suppression is less pronounced for the conformal coupling
\begin{equation}
\sigma_{\textrm{scat}}\stackrel{x\ll 1}{\simeq}\frac{m_f^6}{2\pi\,M^8}\times\left\{\begin{array}{cc}
x & \textrm{conformal},\\
x^4 & \textrm{disformal}.
\end{array}\right. 
\label{eq:sigmascatnonrel}
\end{equation}
In the high energy limit, characterized by $s\gg m_f^2$, both scattering cross sections are of similar order
\begin{equation}
\sigma_{\textrm{scat}}\stackrel{s\gg m_f^2}{\simeq}\frac{s^3}{A\,\pi\,M^8},
\label{eq:sigmascathighenergy}
\end{equation}
where $A = 32$ ($90$) in the conformal (disformal) case. 

In particular, assuming $M_{\textrm{SM}}\sim \textrm{TeV}$ and treating the proton as a Dirac fermion with mass $m_f=m_p\simeq 1~\textrm{GeV}$, we obtain a tiny scattering cross section of energetic $\phi$ off the proton at rest, which is of order $\sigma_{\textrm{scat}}\sim (0.01~\textrm{ab})\times(E_\phi/1~\textrm{TeV})^3$. The cross section, however, can grow to larger values with increasing $E_\phi$, at least up until the validity of the EFT approach is maintained.

For completeness, we also provide the expression for ultra-light disformally coupled scalars scattering off complex scalar DM particles with mass $m_\chi$
\begin{equation}
\sigma_{\textrm{scat,D,c.scal.}} = \frac{s^3}{384\pi\,M_D^8}\,\left(1-\frac{m_\chi^2}{s}\right)^3\,\left[1+\frac{2m_\chi^2}{s} - \frac{m_\chi^4}{s^2}\right].
\end{equation}
This expression also vanishes in the non-relativistic limit, $\sigma_{\textrm{scat,c.scal.}}\sim x^3$, where $x = E_\phi/m_\chi$.

\section{Modified gravity and gravitational waves\label{app:EMgravwaves}}
The theory we considered in the paper is written in the Einstein frame as 
\begin{eqnarray}
    {\cal S} &=& \int d^4 x \sqrt{-g} \left[ {\cal L}_{\rm EH} + \frac{1}{2}g^{\mu\nu}\phi_{,\mu}\phi_{,\nu} - V(\phi)\right]  \nonumber \\
    & &+ {\cal S}_{\rm SM} + {\cal S}_{\rm DM}~,
\end{eqnarray}
where ${\cal S}_{\rm SM}$ and ${\cal S}_{\rm DM}$ are the action for the standard model and dark matter sectors, respectively. These sectors depend on the metrics ${\tilde g}_{\rm SM}$ and  ${\tilde g}_{\rm DM}$, which are related by a disformal transformation to the metric $g$, cf. \cref{eq:metric}. We can write this action in terms of the metric ${\tilde g}_{\rm SM}$ instead by employing a disformal transformation. As a result, the standard model sector is now decoupled from the scalar field $\phi$, but the gravitational sector is no longer of the  Einstein--Hilbert form. Instead, it will in general be of the Horndeski form, in which the Lagrangian reads
\begin{equation}
    {\cal L} = \sum_i {\cal L}_i,
\end{equation}
with
\begin{eqnarray}
{\cal L}_2 &=& K(\phi, X) \nonumber \\
{\cal L}_3 &=& G_3(\phi, X)\box \phi \nonumber \\
{\cal L}_4 &=& G_4(\phi, X) R \nonumber \\
&-& G_{4,X}(\phi,X) \left[(\Box \phi)^2 - (\nabla_\mu\nabla_\nu\phi)^2 \right] \nonumber \\
{\cal L}_5 &=& G_5(\phi,X) G_{\mu\nu}\nabla^\mu \nabla^\nu \phi \nonumber \\
&+& \frac{G_{5,X}(\phi,X)}{6} \left( (\Box \phi)^3 - 3(\Box \phi) (\nabla_\mu\nabla_\nu\phi)^2 \right. \nonumber \\
&+& \left. 2 (\nabla_\mu\nabla_\nu \phi)^3 \right)~, \nonumber
\end{eqnarray}
where $X$ is the kinetic term for the scalar field $\phi$. The transformation rules can be found in Ref.~\cite{Bettoni:2013diz}. Under a purely disformal transformation the Einstein--Hilbert action transforms as follows, i.e. we find that in the new frame ($D = 2/M_{\rm SM}^4$)
\begin{eqnarray}
K(\phi, X) &=& (1 + 2XD)^{1/2}X \nonumber \\
G_3 &=& 0 = G_5\nonumber \\
G_{4} &=& (1 + 2XD)^{1/2} \nonumber
\end{eqnarray}
In general, the speed of gravitational waves $c_T$ depends on the following combinations~\cite{Kobayashi:2011nu}
\begin{eqnarray}
{\cal F}_T &=& 2G_4 + XG_{5,\phi} - 2X\ddot\phi G_{5,X} \nonumber \\
{\cal G}_T &=& 2G_4 - 4X G_{4,X} - XG_{5,\phi} - 2 H X \dot\phi G_{5,X} \nonumber
\end{eqnarray}
with $c_T^2 = {\cal F}_T/{\cal G}_T$. In our case we find 
\begin{equation}
c_T^2 = 1 + 2XD = 1 + 4X/M^{4}_{\rm SM},
\end{equation}
which leads to \cref{eq:cT} for negligible spatial field fluctuations in the vicinity of heavy objects. 

\bibliography{main}

\end{document}